\documentclass{PoS}

\PoS{PoS(LAT2005)341}

\usepackage{epsfig}
\usepackage{amsmath,epsfig,amssymb,eufrak,dsfont}
\usepackage{amscd,latexsym}

\def\lsi{\raise0.3ex\hbox{$<$\kern-0.75em\raise-1.1ex\hbox{$\sim$}}}
\def\gsi{\raise0.3ex\hbox{$>$\kern-0.75em\raise-1.1ex\hbox{$\sim$}}}

\input macros.sty      
\input eqalign.sty     

\title{3-point functions from twisted mass lattice QCD at small quark masses
}

\ShortTitle{Moments from Twisted Mass}

\author{Stefano Capitani\\
        Institut f\"ur Physik/Theoretische Physik\\
        Universit\"at Graz, A-8010 Graz, Austria\\
        E-mail: \email{stefano.capitani@uni-graz.at}}

\author{\speaker{Karl Jansen}, Mauro Papinutto, Andrea Shindler, Ines Wetzorke\\
        John von Neumann-Institut f\"ur Computing NIC, \\             
        DESY, Platanenallee 6, D-15738 Zeuthen, Germany\\
E-mail: \email{Karl.Jansen,Mauro.Papinutto,Andrea.Shindler,Ines.Wetzorke@desy.de}}

\author{Carsten Urbach\\
        Institut f\"{u}r Theoretische Physik, Freie Universit\"{a}t Berlin\\
        Arnimallee 14, D-14195 Berlin, Germany\\
        E-mail: \email{Carsten.Urbach@physik.fu-berlin.de}}

\abstract{
We show at the example of the
matrix element between pion states of a twist-2, non-singlet operator that
Wilson twisted mass fermions allow to compute this phenomenologically
relevant quantitiy at small pseudo scalar masses of O(270 MeV).
In the quenched approximation, we investigate the scaling behaviour of
this observable that is derived from a 3-point function
by applying two definitions of the
critical mass and find a scaling compatible with the expected O($a^2$)
behaviour in both cases. A combined continuum extrapolations allows
to obtain reliable results at small pion
masses, which previously could not be explored by lattice QCD simulations.}

\FullConference{XXIIIrd International Symposium on Lattice Field Theory\\
                 25-30 July 2005\\
                 Trinity College, Dublin, Ireland}

\begin{document}

\section{Introduction}

Precision experiments allow to determine the parton distribution functions
$f(x,Q_0^2)$ as a function of the momentum fraction $x$ and the 
energy scale $Q_0^2$. At a fixed value of $Q_0^2$, 
this function is usually parametrized as 
$f(x)\propto x^\alpha(1-x)^\beta$ with exponents $\alpha$, $\beta$ that
are obtained from global fits of the world-wide available 
experimental data (see ref.~\cite{Devenish:2004pb} for a general introduction). 
The bridge to results from non-perturbative calculations in lattice QCD 
are the moments $\int dx x^{n-1} f(x)$ of the parton distribution 
functions which are directly computable 
with lattice technues. 
From the phenomenological side, 
there has been a lot of work to estimate the errors
for these moments \cite{Blumlein:2002be,Alekhin:2002fv,Thorne:2004ci}. 
These investigations also demonstrated clearly that 
the uncertainties of the parton distribution functions can easily 
reach 100\%, in particular 
for the gluon distribution function. 
Hence, results from lattice calculations are highly welcome 
for phenomenologists and experimentalists. 

Lattice calculations for moments of parton distribution functions have 
progressed very much over the last years. We control the continuum limit 
\cite{Guagnelli:1999gu}
and the finite size effects \cite{Guagnelli:2004ww},
we understand fully the non-perturbative, scale dependent renormalization 
\cite{Guagnelli:2003hw}
of the operators needed to compute the moments and we can 
obtain matrix elements of such operators with a good precision.
Nevertheless, even in the quenched approximation, there remains an 
important and hard to quantify systematic
error: so far simulation data are obtained at only rather large values of the 
pseudo scalar mass of 600MeV of even higher. Clearly, an extrapolation to the 
physical value, i.e. a pion mass of about 140MeV, is difficult since
contact to chiral perturbation theory cannot be established with such 
large pseudo scalar masses. 

In fig.~9 of ref.~\cite{Guagnelli:2004ga} we give an example of the 
situation of our earlier 
calculation  
of a twist-2, non-singlet operator in pion states, 
correponding to the average momentum of the parton. The gap between
the smallest pseudo scalar mass of about 600MeV reached in this work 
and the physical point 
at 140MeV
appears to be quite large, leaving a substantial uncertainty in confronting
the lattice calculations with the experimental results. 


In this contribution we will report about a way to overcome the 
uncertainty of the chiral extrapolation by employing Wilson twisted mass
lattice fermions 
\cite{Frezzotti:2000nk,Frezzotti:2003ni}
which allows to regulate unphysical, small eigenvalues of the Wilson-Dirac 
operator, and hence can 
reach pion masses of well below 
300MeV and thus provide a tool to use chiral perturbation theory to
perform the last, small step to extrapolate to the physical point.
See \cite{Frezzotti:2004pc,shindler} for recent reviews.

We emphasize that it is the main goal of this contribution to show that the 
gap in the pseudo scalar masses, seen in fig.~9 of 
ref.~\cite{Guagnelli:2004ga}, can indeed be
closed when using Wilson twisted mass fermions. It is not the aim of the 
paper to provide a phenomenological number of the matrix element under
consideration here and to compare to experiment. 
At the example of a 3-point function, the paper shall provide a 
demonstration that 
the Wilson twisted mass approach can be used to simulate 
small pion masses of O(270 MeV) while keeping O($a^2$) cut-off effects
under control as has been shown already for quantities that derive 
from 2-point correlation functions 
\cite{Abdel-Rehim:2005gz,Jansen:2005gf,Jansen:2005kk}.  
A more detailed account of this work will be published in ref.~\cite{Capitaniprep}.

\section{Lattice action and operators}

In this paper we will work on a lattice $L^3 \times T$ 
with twisted mass fermions the action of which can be written as
\begin{equation}
  \label{tmaction}
  S[U,\psi,\bar\psi] = a^4 \sum_x \bar\psi(x) ( D_W + m_0 + i \mu
\gamma_5\tau_3 ) \psi(x)\; ,
\end{equation}
with the standard Wilson-Dirac operator $D_{\rm W}$
and $m_0$, $\mu$ the bare un-twisted and twisted mass parameters, 
respectively. 
Here and in the following $\psi(x)$ indicates a flavour doublet of quarks.  

In the present work, we will employ two 
definitions for the critical mass with the final aim of a combined
continuum extrapolation of the results obtained in the two cases. 
The first definition of the critical mass is the point where 
the pseudoscalar meson mass vanishes, the second, where the 
PCAC quark mass vanishes. In the following we will refer to the first
situation as the ``pion definition'' and to the second situation 
as the ``PCAC definition'' of the critical point,
as introduced already in \cite{Jansen:2005kk}.
Both definitions should lead to $O(a)$-improvement
\cite{Frezzotti:2003ni,Aoki:2004ta,Sharpe:2004ps,Sharpe:2004ny}, but 
they can induce very different $O(a^2)$ effects, in particular at 
small pseudoscalar meson masses \cite{Frezzotti:2005gi}. 

The towers of twist-2 operators related to the unpolarized structure functions
have the following expressions
\be
O^a_{\mu_1 \cdots \mu_N}(x) = \frac{1}{2^{N-1}} \bar\psi(x)\gamma_{\{\mu_1}
\lrD_{\mu_2}\cdots \lrD_{\mu_N\}}\frac{1}{2} \tau^a\psi (x)\; ,
\label{eq:op}
\ee
where $\{\cdots\}$ means symmetrization on the Lorentz indices,
$
\lrD_{\mu} = \rD_{\mu} - \lD_{\mu}$ and $\qquad  D_{\mu} =
\frac{1}{2}[\nabla_\mu + \nabla_\mu^{*}]\; .
$
The flavour structure is specified by the Pauli matrices $\tau^a$ where we
include here also the identity with $\tau^4=2 \cdot \mathds{1}$.   
In general one should perform an axial rotation in order to have the
expressions for the twist-2 operators for the twisted mass formulation.
However, the operators in
eq. (\ref{eq:op}) with flavour index $a=1,4$ do not rotate.
We concentrate in this
work on the twist-2 quark operator related to the lowest moment 
of the valence quark parton distribution function in a pion.
In particular we will use for the {\it up} quark (the {\it down} quark can be treated in
the same way)  
%
\be
\cO^u_{44}(x) = \frac{1}{2} \bar\psi(x) \Big[ \gamma_4 \lrD_4 - {1 \over 3}
\sum_{k=1}^3 \gamma_k \lrD_k \Big] \frac{(1+\tau^3)}{2} \psi(x)\; .
\label{O44}
\ee

The matrix elements of this operator can be computed in the standard way,
described in refs. \cite{Martinelli:1987zd,Best:1997qp}. We indicate with 
$
P^{\pm}(x) = \bar\psi(x) \gamma_5 \frac{\tau^{\pm}}{2} \psi(x)$ and      
$\tau^{\pm} = \frac{\tau^1 \pm i\tau^2}{2}
$
the interpolating operator for the charged pion.
The ratio of the 2-point and 3-point functions
\be
C_P(x_4) = a^3\sum_{{\bf x}}\langle P^+(0) P^-({\bf x},x_4) \rangle\; ,\;\;
C_{44}(y_4) = a^6 \sum_{{\bf x},{\bf y}}\langle P^+(0) \cO_{44}({\bf y},y_4)
P^-({\bf x},T/2) \rangle\; ,
\ee
is related to the matrix element we are interested in.
In particular if we perform a transfer matrix decomposition and we define
$
R(y_4) = \frac{C_{44}(y_4)}{C_P(T/2)}\; ,
$
we obtain, in the limit when only the fundamental state dominates ($0 \ll y_4
\ll T/2$), 
the relevant bare quantity as  
\be
\langle x \rangle^{\rm bare} = \frac{1}{m_\pi}\cdot R\; .
\ee

The matrix element obtained in this way has to be renormalized by a 
multiplicative renormalization factor. 
To this end, we took the $Z_\mathrm{RGI}$ as computed in 
ref.~\cite{Guagnelli:2004ga,Guagnelli:2003hw}
using the Wilson action.
The renormalized matrix element has a well defined continuum limit and in the
phenomenological relevant $\msbar$ scheme is given by 
\be
\langle x \rangle^{\msbar} (\mu,r_0m_\pi) = \lim_{a\to 0}
\frac{\langle x \rangle^{\rm bare}(a,m_\pi) Z_\mathrm{RGI}(a) }{f^{\msbar}(\mu)} \; \qquad
\mu=2GeV ,
\ee
where $Z_\mathrm{RGI}$ and $f^{\msbar}(\mu)$ were computed in
ref. \cite{Guagnelli:2004ga} (see this reference for further details).

\section{Numerical results}

Our quenched simulations were performed for a number of bare quark masses in a
corresponding pion mass range of $270 \mathrm{~MeV} < m_\pi < 1.2 \mathrm{~GeV}$
using the Wilson plaquette gauge action, employing 
periodic boundary conditions for all fields.
In ref.~\cite{Capitaniprep} detailed tables with the numerical results can be found.
We performed simulations at fixed values of the twisted mass parameter 
$\mu$ while keeping $m_0$ to be the critical quark mass as obtained 
from the pion or PCAC definitions. 

Using the force parameter $r_0=0.5 \rm{~fm}$
to set the scale we determined the
physical value of the pion mass and interpolated the values of 
$\langle x \rangle$ for the chosen values of $m_{\rm PS} r_0$. 
These values are close to the simulated ones, and so even a linear
interpolation is usually sufficient. The lowest mass that can be reached 
corresponds to $m_{\rm PS}=272$ MeV.
Note that we corrected our values for $\langle x \rangle$ for finite size effects using the 
results of ref.~\cite{Guagnelli:2004ww}.

\begin{figure}[htb]
\vspace{-0.0cm}
\begin{center}
\epsfig{file=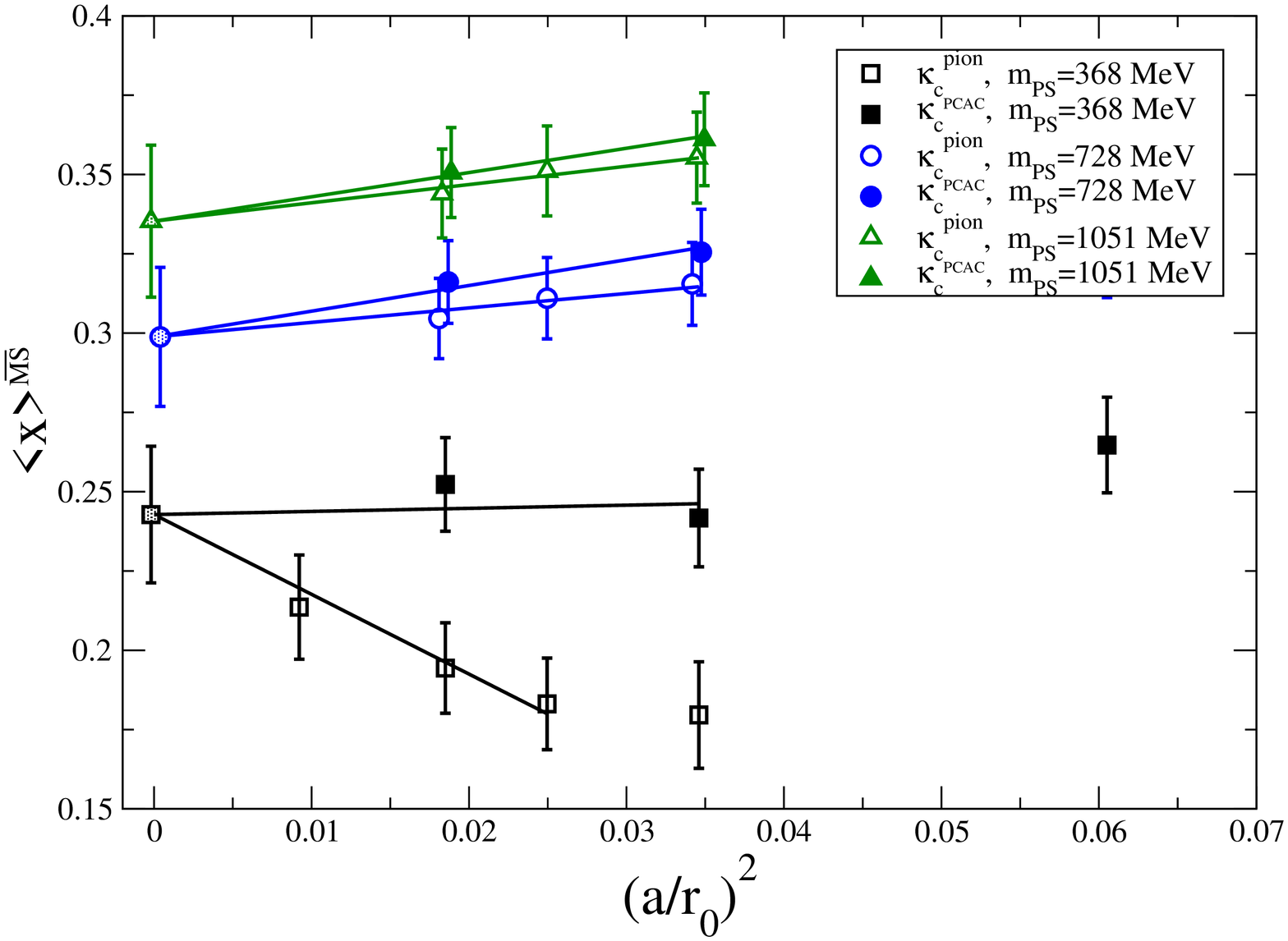,angle=270,width=0.49\linewidth}
\epsfig{file=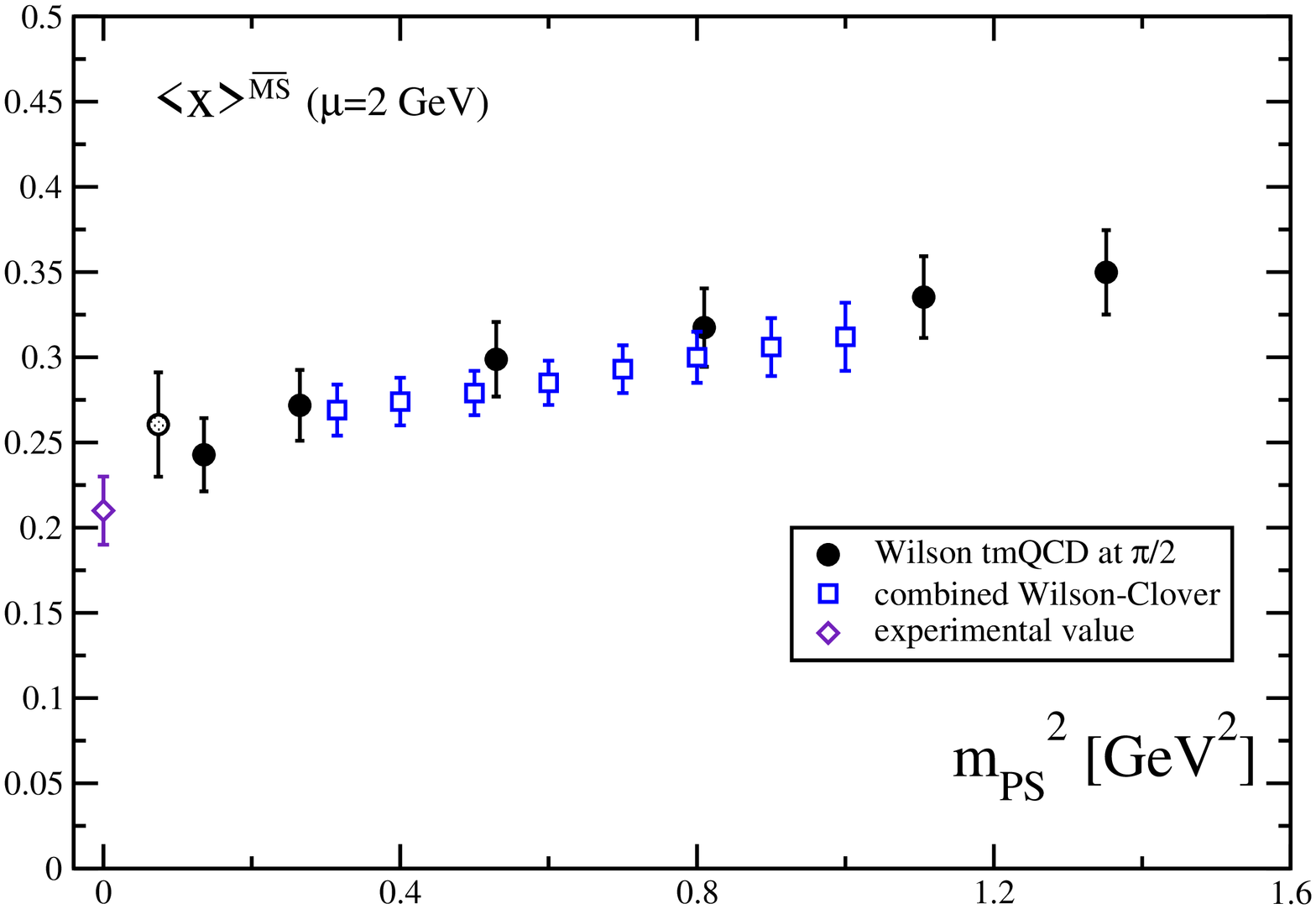,angle=270,width=0.49\linewidth}
\end{center}
\vspace{-0.0cm}
\caption{Left panel: $\langle x \rangle^{\msbar}$ as a function of $a^2$ for different
  values of the pion mass.  
Right panel:
$\langle x \rangle^{\msbar}(\mu= \ 2 \ GeV)$ extrapolated to the continuum as a function of
the pion mass. Open squares represent results that are obtained from a 
combined continuum
extrapolation of earlier Wilson and clover-Wilson simulations,
ref.~\cite{Guagnelli:2004ga}. The filled cicles represent our results using
Wilson twisted mass fermions. The open circle denotes 
a result which is not corrected for finite size effects and the diamond corresponds 
to the experimental point.
\label{fig:me}}
\end{figure}

In fig.~\ref{fig:me} we show the continuum extrapolation of 
$\langle x\rangle$, already converted to the $\msbar$ scheme as explained
in the previous section, at fixed pion mass for a wide range of pion masses. 
The scaling of $\langle x\rangle$ is in agreement with the expected 
O($a^2$) cut-off effects for both definitions of the critical mass.
The slope of $\langle x\rangle$ as a function of $a^2$ appears to be
rather small, at least for the pion masses investigated here. It also
seems that the slope changes sign from positive to negative values
when the pion mass is decreased.
The plot also reveals that the continuum extrapolations for $\langle x \rangle$ performed 
independently for both definitions of the critical mass nicely
agree. 


In fig.~\ref{fig:me} (right panel) we show 
$\langle x \rangle^{\msbar}$ in the continuum as a function of the continuum
pion mass. The open squares represent our
values obtained earlier from a combined continuum extrapolation of Wilson
and clover-improved Wilson fermions using the Schr\"odinger functional scheme
\cite{Guagnelli:2004ga}. As usual, such simulations have to stop at a pion mass
of about 600 MeV. At such high pion masses it becomes very difficult, if not
impossible to compare the simulation results, even when extrapolated to 
the continuum limit as done here, to chiral perturbation theory. 

The right panel of Fig.~\ref{fig:me} shows that 
with Wilson twisted mass fermions,  represented by the filled 
circles, the large gap between a pion mass of 
about 600 MeV, as the lower bound for standard simulations, and the physical
value can be bridged. The Wilson twisted mass results for 
$\langle x \rangle^{\msbar}$ are consistent with a linear behaviour. 
Note that we add in the figure a point denoted with an open circle which 
is not corrected for finite size effects and is shown for illustration
purpose only that also pion mass below 300MeV can be reached with Wilson
twisted mass fermions.


\section{Conclusions}

In this contribution we have computed the lowest moment 
of a pion parton distribution funtion. In particular, we concentrated
on a twist-2, non-singlet operator the matrix element of which 
corresponds to the average momentum of a parton in the pion. 
We demonstrated, that also for such matrix elements pseudo scalar masses 
below 300MeV can be reached.  
At the same time, the cut-off effects even at such small pseudo scalar
masses appear only in O(a$^2$) and remain small and controllable. 
This clearly opens the door to make contact to chiral perturbation theory
and to perform reliable extrapolations to the physical point with a 
pseudo scalar mass that corresponds to the experimentally observed 
pion mass. 
 
In the present work we have employed two definitions of the critical
mass, the pion and PCAC mass definition \cite{Jansen:2005gf,Jansen:2005kk}. 
We could perform a controlled continuum extrapolation of $\langle x\rangle$
at least down to pion masses of about 270 MeV by combining the data obtained
with the two definitions.  
Of course, in principle, also overlap simulations are able to reach such
values of the pion mass. This will come, however, at a much higher 
simulation cost \cite{Chiarappa:2004ry}.
The right panel of our final figure, fig.~\ref{fig:me}, 
clearly demonstrates that with our
present setup it is possible to go beyond standard calculations and bridge the
gap between large pion mass values of 600 MeV and the physical value of the pion
mass.

\section{Acknowledgements}
We thank Giancarlo Rossi and Stefan Sint                                                     
for many useful discussions. 
The computer centers at NIC/DESY Zeuthen, NIC at Forschungszentrum
J{\"u}lich and HLRN provided the necessary technical help and computer
resources. 
S.~C.~gratefully acknowledges support by Fonds zur F\"orderung der
Wissenschaftlichen Forschung in \"Osterreich, Project P16310-N08.
This work was supported by the DFG 
Sonderforschungsbereich/Transregio SFB/TR9-03.

\bibliographystyle{JHEP-2}
\bibliography{3ptpos}

\end{document}